\documentclass[aps,pra,twocolumn,amsmath,amssymb,nofootinbib,superscriptaddress]{revtex4-2}
\usepackage[colorlinks,
           linkcolor=blue,
            anchorcolor=blue,
            allcolors = blue,
            citecolor=blue
            ]{hyperref}

\usepackage{bm}
\usepackage{graphicx}
\usepackage{enumerate}

\begin{document}


\title{Simultaneous determination of multiple low-energy eigenstates of many-body systems on a superconducting quantum processor}

\author{Huili Zhang}
\affiliation{Beijing Key Laboratory of Fault-Tolerant Quantum Computing, Beijing Academy of Quantum Information Sciences, Beijing
100193, China}

\author{Yibin Guo}    
\affiliation{Beijing Key Laboratory of Fault-Tolerant Quantum Computing, Beijing Academy of Quantum Information Sciences, Beijing
100193, China}
\affiliation{Institute of Physics, Chinese Academy of Sciences, Beijing 100190, China}
\affiliation{University of Chinese Academy of Sciences, Beijing 101408, China}

\author{Guanglei Xu}
\affiliation{Institute of Physics, Chinese Academy of Sciences, Beijing 100190, China}

\author{Yulong Feng}
\affiliation{Beijing Key Laboratory of Fault-Tolerant Quantum Computing, Beijing Academy of Quantum Information Sciences, Beijing
100193, China}

\author{Jingning Zhang}
   \email{zhangjn@baqis.ac.cn}
\affiliation{Beijing Key Laboratory of Fault-Tolerant Quantum Computing, Beijing Academy of Quantum Information Sciences, Beijing
100193, China}

\author{Hai-feng Yu}
    \email{hfyu@baqis.ac.cn}
\affiliation{Beijing Key Laboratory of Fault-Tolerant Quantum Computing, Beijing Academy of Quantum Information Sciences, Beijing
100193, China}
\affiliation{Hefei National Laboratory, Hefei 230088, China}

\author{S. P. Zhao}
\affiliation{Beijing Key Laboratory of Fault-Tolerant Quantum Computing, Beijing Academy of Quantum Information Sciences, Beijing
100193, China}
\affiliation{Institute of Physics, Chinese Academy of Sciences, Beijing 100190, China}


\begin{abstract}
The determination of the ground and low-lying excited states is critical in many studies of quantum chemistry and condensed-matter physics. Recent theoretical work proposes a variational quantum eigensolver using ancillary qubits to generate entanglement in the variational circuits, which avoids complex ansatz circuits and successive measurements in the previous algorithms. In this work, we employ the ancilla-entangled variational quantum eigensolver to simultaneously compute multiple low-lying eigenenergies and eigenstates of the $\text{H}_2$ molecule and three- and five-spin transverse field Ising models (TFIMs) on a superconducting quantum processor. We obtain the potential energy curves of $\text{H}_2$ and show an indication of antiferromagnetic to paramagnetic phase transition in the TFIMs from the average absolute magnetization. Our experiments demonstrate that the algorithm is capable of simultaneously determining multiple eigenenergies and eigenstates of many-body systems with high efficiency and accuracy and with less computational resources.
\end{abstract}

\maketitle

\section{Introduction}

\begin{figure*}
    \centering
    \includegraphics[width=0.99\textwidth]{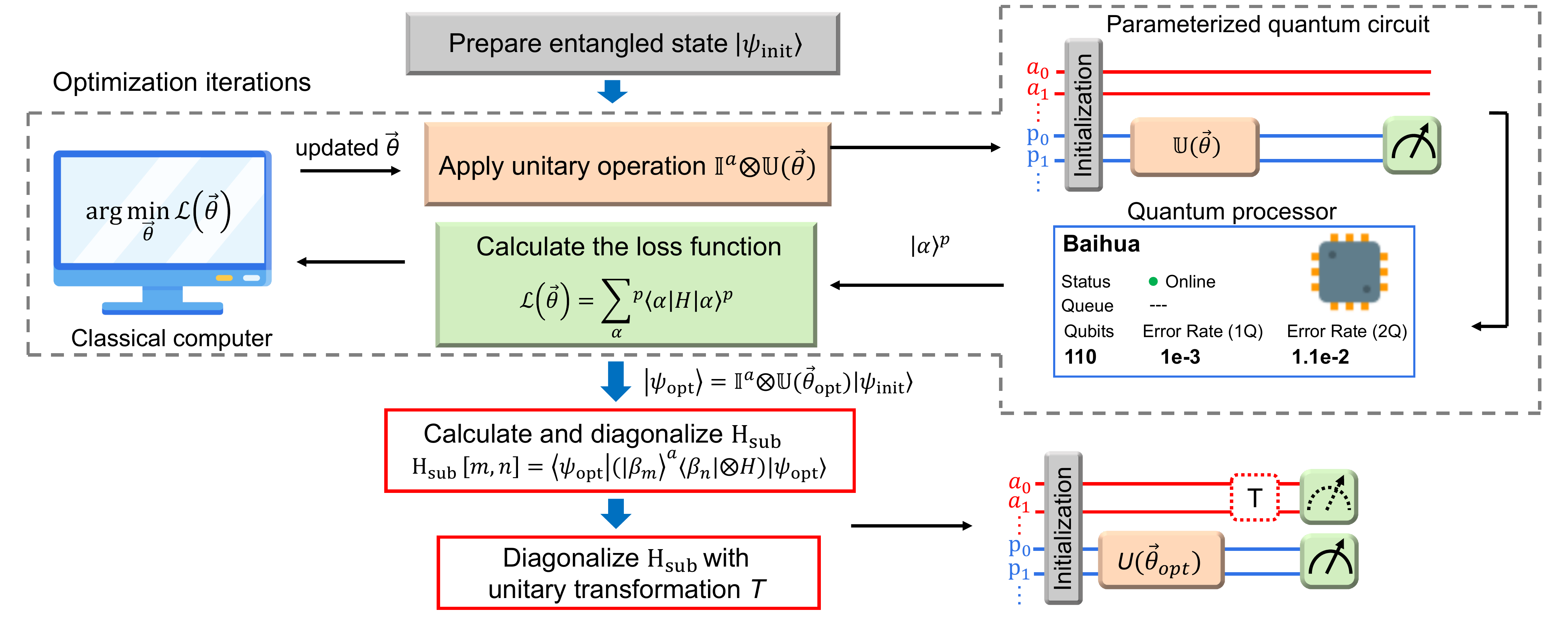}
    \caption{Schematic diagram of AEVQE. The quantum circuit consists of ancillary qubits $a_i$ and physical qubits $p_i$. First, the ancillary and physical qubits are initialized to entangled states. In the optimization iterations, the variational circuits with parameters $\vec{\theta}$ are executed on the quantum processor. The measured results of the physical qubits (noted as $|\alpha\rangle^{p}$)  are fed into the classical computer for searching $\vec{\theta}_{\text{opt}} = \text{arg}\mathop{\text{min}}\limits_{\vec{\theta}}\mathcal{L}(\vec{\theta})$. After optimization, $\vec{\theta}_{\text{opt}}$ is obtained and applied to calculate $H_{\text{sub}}$. Finally, unitary transformation $T$ is applied to diagonalize $H_{\text{sub}}$ and determine the eigenenergies and eigenstates.}   
    \label{Fig1}
\end{figure*}

Quantum computers have potential computational advantages over classical computers~\cite{lloyd1996universal, georgescu2014quantum, arute2019quantum, zhong2020quantum, madsen2022quantum} in various applications such as integer factoring \cite{shor1999polynomial}, quantum simulation \cite{aspuru2005simulated, yuan2020quantum, mcardle2020quantum, bauer2023quantum, gerritsma2010quantum, bloch2012quantum}, and quantum artificial intelligence \cite{jordan2015machine, carleo2019machine, deng2017machine, bausch2024learning, alexeev2024artificial, koutromanos2024control}.  Variational quantum algorithms (VQAs) are one of the leading algorithms for achieving quantum advantage in current noisy intermediate-scale quantum (NISQ) devices~\cite{cerezo2021variational,bharti2022noisy,kandala2017hardware, sweke2020stochastic, tang2021qubit}. The first VQA, the variational quantum eigensolver (VQE), is developed to find the ground state of a specific Hamiltonian $H$~\cite{google2020hartree, peruzzo2014variational, liu2019variational, nam2020ground, hempel2018quantum,wang2023electronic}, which utilizes a parameterized ansatz quantum circuit $U(\theta)$ to generate an ansatz state $|{\psi(\theta)}\rangle$. By minimizing the expectation value $\langle{\psi(\theta)}|H|{\psi(\theta)}\rangle$ (i.e., the loss function) via iterative optimization of the circuit parameters $\theta$, the ground state of the target Hamiltonian can be found. The parameterized  quantum circuit in the algorithm enables the reduction of the quantum circuit depth, which is the key to realize quantum advantage for NISQ devices.

Recently, the VQEs have been extended to calculate the low-energy excited states~\cite{ibe2022calculating, higgott2019variational, jones2019variational, benavides2024quantum, smart2024many, jouzdani2021method, nakanishi2019subspace, han2024multilevel, yalouz2021state}, which play a crucial role in various studies like the prediction and analysis of chemical reactions and phase transitions. To do so, the orthogonality of the calculated eigenstates is of primary importance. In the variational quantum deflation (VQD) algorithm, the $k$ eigenvalues $E_{0},...,E_{k} $ are computed recursively from lowest to highest, with the loss function given by $\mathcal{L(\theta)} = \langle\psi_{k}|H|\psi_{k}\rangle + \sum_{i=0}^{k-1}\beta_{i}|\langle\psi_k|\psi_i\rangle|^{2}$, where $\beta_{i}$ are chosen to be sufficiently large to ensure the orthogonality of the eigenstates~\cite{ibe2022calculating, higgott2019variational, jones2019variational, benavides2024quantum, smart2024many, jouzdani2021method}. In the subspace-search VQE (SSVQE) method, $k$ orthogonal initial states $|\psi _{i}\rangle$ are prepared and evolved to $U(\theta)|\psi_{i}\rangle$ through the parameterized circuit, and the orthogonality of the output states is ensured by the unitarity of $U(\theta)$. The loss function with weighted sum of the $k$ orthogonal output states in the form $\mathcal{L}(\theta) = \sum_{i=0}^{k}w_{i}\langle\psi_{i}|U^{\dagger}(\theta)HU(\theta)|\psi_{i}\rangle$ is used~\cite{nakanishi2019subspace, han2024multilevel, yalouz2021state}. 

While these methods have demonstrated respective achievements, they also show shortcomings in the calculations of the multiple low-lying excited states. In VQD, the accumulative energy errors from the low-lying states may reduce the accuracy of the high eigenenergies~\cite{ibe2022calculating}. In SSVQE, multiple initialization and variational circuits are required, thus increasing the depth of the quantum circuit and the number of optimization parameters~\cite{nakanishi2019subspace, han2024multilevel}. To overcome these difficulties, a VQE algorithm capable of \textit{simultaneously} determining multiple eigenstates have been proposed \cite{xu2023concurrent}. The method uses a set of ancillary qubits to construct maximally entangled states with corresponding qubits in the system, so that multiple final eigenstates evolved from the initial orthogonal states can be obtained simultaneously from different ancillary states. 

In this work, we present the experimental demonstration of the VQE algorithm with the superconducting quantum processor \textit{Baihua} on the platform \textit{Quark}~\cite{QuafuSQC}, which is referred to as ancilla-entangled VQE (AEVQE). We apply the algorithm to simulate the $\text{H}_{2}$ molecule and transverse field Ising models (TFIMs). For the $\text{H}_{2}$ molecule, we obtain the H-H bond distance dependence of two eigenenergies, with an average energy error of 0.027 Hatree for the first excited state. For the TFIMs, combined with symmetry verification methods, we calculate four eigenenergies for the three-spin system and two eigenenergies for the five-spin system, with average errors of of 0.029 and 0.099 for the high-lying excited states, respectively. Moreover, we show an indication of the phase transition of the model by calculating the average absolute magnetization.  Our results demonstrate the capability of AEVQE for determining multiple eigenenergies and eigenstates of many-body systems on quantum hardwares. 

\section{Scheme} 

To determine $K$ eigenenergies and eigenstates of $H$ (embedded in $N_{p}$ physical qubits), additional $N_{a}$ ancillary qubits $(\log_{2}K\leq N_{a} < N_{p})$ are required. The process of AEVQE is illustrated in Fig. \ref{Fig1} and is described as follows \cite{xu2023concurrent}: 
\begin{enumerate}[(1)]
\setlength{\itemindent}{1.1em}
\setlength{\itemsep}{0pt}
\item Choose $N_{a}$ physical qubits and prepare an entangled state between the $i$th ancillary qubit and $i$th physical qubit to construct an initial state $|\psi_{\text{init}}\rangle$.
\item Apply a unitary operation $\mathbb{I}^{a} \mathop{\otimes}\mathbb{U}(\vec{\theta})$ to the initialized system, where $\mathbb{I}^{a}$ represents the identity operators on ancillary qubits, and $\mathbb{U}(\vec{\theta})$ represents the unitary operators on physical qubits. $\vec{\theta}$ is the variational parameter. 
\item Measure the physical qubits and calculate the loss function
\begin{equation}
    \mathcal{L(\vec{\theta})} = \sum_{\alpha =0}^{K-1}\ ^{p}\langle\alpha|H|\alpha\rangle^{p} \geq \frac{1}{K}\sum\limits_{i=0}^{K-1}E_{i},
\end{equation}

where $|\alpha\rangle^{p} $ is the final state of the physical qubits. The loss function sets an upper limit on the average $K$ eigenenergies of $H$.

\item After the convergence of the loss function, measure the ancillary qubits and the physical qubits to calculate $H_\text{sub}$. The matrix elements of the  $H_{\text{sub}}$ are given by 
\begin{equation}
    H_{\text{sub}}[m,n] = \langle\psi_{\text{opt}}|(|\beta_m\rangle^{a} \langle\beta_n|\otimes H)|\psi_{\text{opt}} \rangle,
\end{equation}
where $|\beta_{m}\rangle^{a}$,  $|\beta_{n}\rangle^{a}$ represent the computational basis of the ancillary qubits, $0\leq m,n\leq K-1$. $|\psi_{\text{opt}}\rangle = \mathbb{I}^{a}\otimes\mathbb{U}(\vec{\theta}_{\text{opt}})|\psi_{\text{init}}\rangle$ is the final state.
\item Diagonalize $H_{\text{sub}}$ with the unitary transformation $T$ on the ancillary qubits to obtain eigenerengies $E_{i}$ and eigenstates $|E_{i}\rangle^{p}$. 
\end{enumerate}
 
The minimization of the loss function is performed by a classical optimizer of the simultaneous perturbation stochastic approximation (SPSA)\cite{spall1992multivariate,kandala2017hardware}. The initial parameters are randomly chosen, at the $k$th iteration step, and the energy gradient $\bm{g}_{k}(\vec{\theta}_{k})$ is calculated by adding the perturbation $\pm\vec{\epsilon}_{k}$ of the input parameter $\vec{\theta}_{k}$ as
\begin{equation}
    \bm{g}_{k}(\vec{\theta}_{k}) = \frac{\mathcal{L} (\vec{\theta}_{k}+\vec{\epsilon}_{k}) - \mathcal{L}(\vec{\theta}_{k}- \vec{\epsilon}_{k})}{2\vec{\epsilon}_{k}}~.
\end{equation}
Then the parameter is updated by the formula $\vec{\theta}_{k+1} = \vec{\theta}_{k} + \eta_{k}\bm{g}_{k}(\vec{\theta}_{k})$, where $\eta_{k}$ is the learning rate.  In this work, we set the perturbation $\vec{\epsilon}_{k} = 0.1$ and the learning rate $\eta_{k} = 0.2$.  

Our experiment is carried out with processor \textit{Baihua} on the platform \textit{Quark} (see Fig.~\ref{Fig1})~\cite{QuafuSQC}. For the measurement errors, we apply readout error mitigation to the classical measurement outcomes. To reduce the sampling errors, the measurements are repeated by 15 $\times$ 1024 times in each iteration. The results are divided into five groups to calculate the loss function. When the variance of the loss function is less than the threshold set for the experiment,  optimization continues. Otherwise, the circuit with the same parameters is executed repeatedly. In cases where the TFIMs are simulated, we apply symmetry verification to reduce gate errors.  These techniques are essential for accurate experimental implementations of AEVQE on the quantum cloud platform. 

\begin{figure}[t]
    \centering
    \includegraphics[width=0.5\textwidth]{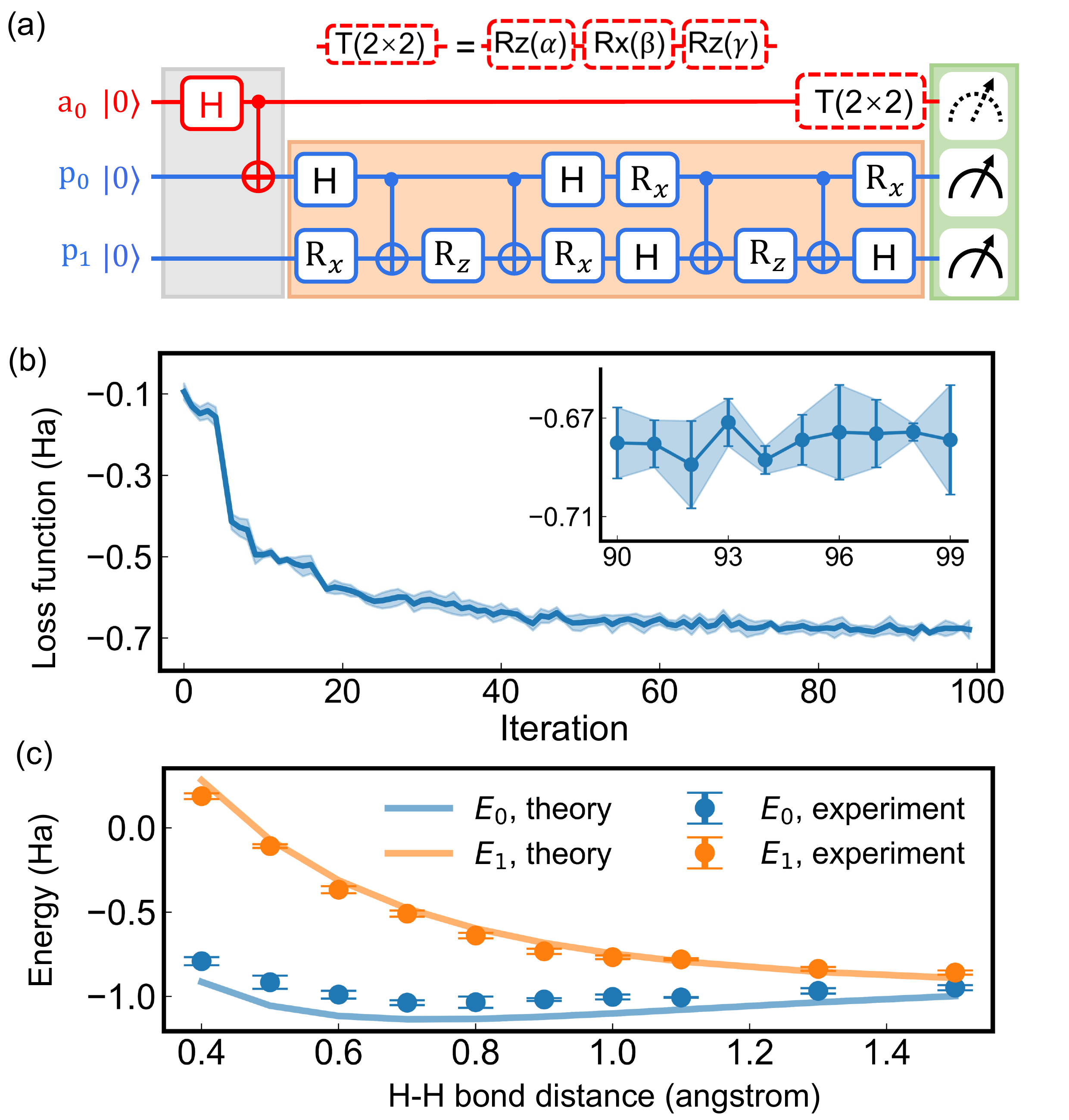}
    \caption{Calculation of the ground and first excited state energies $E_0$ and $E_1$ of the $\text{H}_{2}$ molecule. (a) The circuit schematic. The gray and orange areas represent the initialization and variational circuits, respectively. $H$ represents the Hadmard gate. $R_x$ and $R_z$ are the rotations along the $x$ and $z$ axes of the Bloch sphere, with the rotation angles to be optimized. The unitary transformation $T$ is applied to diagonalize $H_\text{sub}$ after the loss function convergence. (b) The optimization of the loss function at H-H bond distance $d = 0.6$ angstrom. The inset shows the result of the final ten iterations. (c) The experimental energy potentials $E_0$ (blue dots) and $E_1$ (orange dots) as a function of the H-H bond distance. The lines are the corresponding theoretical results. The error bars represent the standard error of the average energy. All energies are in the unit of Hatree (Ha).}
    \label{Fig2}
\end{figure}  

\section{Simulation of the $\mathbf{H_2}$ molecule}

We first use AEVQE to experimentally calculate the ground and first excited state energies of the $H_2$ molecule. Two physical qubits are required to implement the variational circuit of the unitary coupled-cluster generalized singles and doubles (UCCGSD) ansatz \cite{hong2024Refining}. The Hamiltonian $H$ is given by 
\begin{equation}
H = c_0 + c_1 Z_0 + c_2 X_0 + c_3 Z_0Z_1 + c_4 X_0X_1,
\end{equation}
here the coefficients $c_{i}$ of each observable depend on the H-H bond distance \cite{mcclean2020openfermion,sun2018pyscf}. Introducing one ancillary qubit enables the simultaneous determination of the two lowest eigenenergies $E_0$ and $E_1$. The quantum circuit is shown in Fig. \ref{Fig2}(a). In the initialization step, the ancillary qubit and the first physical qubit are prepared in the Bell state $(|00\rangle+|11\rangle)/\sqrt{2}$, and the second physical qubit is prepared in $|0\rangle$. In the variational step, the unitary circuit is applied to two physical qubits, and the rotation angles of $R_x$ and $R_z$ are variational parameters. In the measurement step, the physical qubits are measured to calculate the loss function. The loss function minimization procedure at the H-H bond distance $d=0.6$ is shown in Fig. \ref{Fig2}(b), which is seen to have a good convergence within 100 iteration steps.  

After the convergence of the loss function, the ancillary qubits and the physical qubits are measured, with the observables for the matrix elements of $H_{\text{sub}}$ shown in Table \ref{table1}. $H_{\text{sub}}$ can be diagonalized with a unitary transformation matrix $T(2\times2)$. In the case of one ancillary qubit, $T(2\times2)$ is decomposed into qubit rotations along $x$ and $z$ axes on the Bloch sphere as illustrated in Fig.~\ref{Fig2}(a).  After diagonalization, the ground state and excited state are prepared in the $|0\rangle$ and $|1\rangle$ subspaces of the ancillary qubit, and eigenenergies and eigenstates can be obtained. By varying the H-H bond distance, we experimentally obtain the potential energy curves $E_0$ and $E_1$, which are shown in Fig. \ref{Fig2}(c). The average errors between experiment and theory are 0.098 Hatree for $E_{0}$ and 0.027 Hatree for $E_1$. The results demonstrate that the calculational accuracy of the excited state eigenenergy is not affected by the errors of the ground state eigenenergy.  

\begin{table}[t]
    \centering
    \caption{The observables for  the matrix elements of $H_{\text{sub}}$ for the calculation of the ground and first excited state energies of $H_2$.}
    \setlength{\tabcolsep}{26pt}
    \begin{tabular}{c|c}
      \toprule
      Observable   &  Matrix element\\\hline
      $(I+Z) \otimes H$    & $H_{\text{sub}}[0,0]$ \\
      $(X+iY) \otimes H$    & $H_{\text{sub}}[0,1]$ \\
      $(X-iY) \otimes H$    & $H_{\text{sub}}[1,0]$ \\
      $(I-Z) \otimes H$    & $H_{\text{sub}}[1,1]$ \\\hline
    \toprule
    \end{tabular}
    
    \label{table1}
\end{table}

\section{Transverse field Ising Model}

\begin{figure*}
    \centering
    \includegraphics[width=0.95\textwidth]{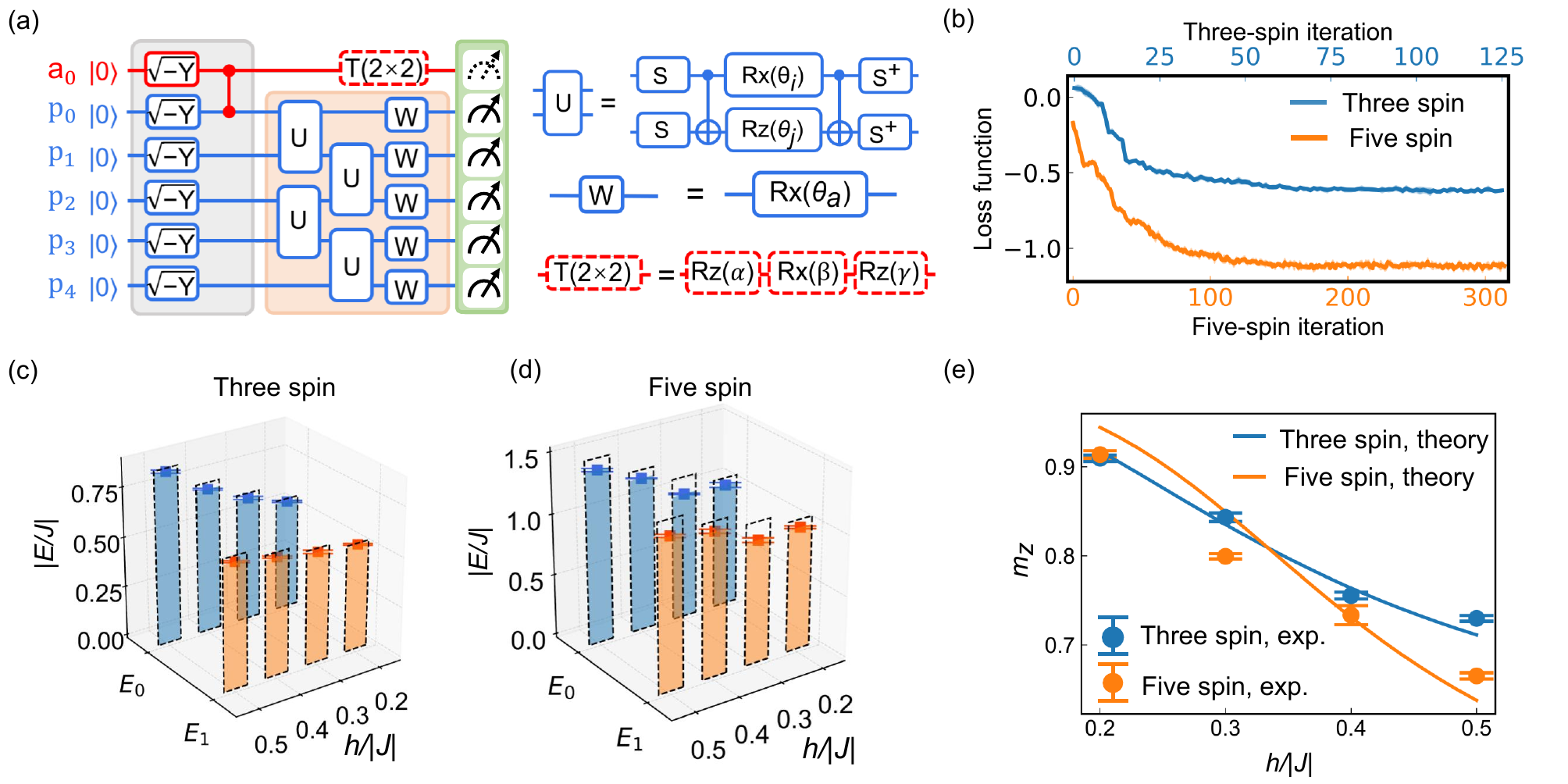}
    \caption{Calculation of the ground and first excited state energies $E_0$ and $E_1$ of the transverse field Ising models with one ancillary qubit. (a) The circuit schematic for the five-spin system.  Decompositions of $U$, $W$, and $T$ are shown on the right. $S = R_z(\pi/2)$ is the phase gate and $\sqrt{-Y} = R_y(-\pi/2)$. (b) The optimization of the loss function for $h/|J| = 0.4$. The blue and orange lines represent the results of the  three- and five-spin TFIMs, respectively. (c) and (d) The experimental ground and first excited state energies $E_0$ and $E_1$ of the three- and five-spin TFIMs. The dashed-line rectangles are the corresponding theoretical results. The error bars represent the standard errors of the average energy. (e) Experimental (dots) and theoretical (lines) average absolute magnetization $m_{\rm z}$ of the three- and five-spin TFIMs.}
    \label{Fig3}
\end{figure*}

The transverse field Ising models are widely studied in diverse fields of many-body quantum physics~\cite{pfeuty1970one, heyl2013dynamical, mondaini2016eigenstate, schmitt2022quantum, li2023probing}. The Hamiltonian of the one-dimensional $N_p$-spin TFIM is described by:
\begin{equation}
    H = -J\sum_{i=0}^{N_p-2} \sigma^{z}_{i}\sigma^{z}_{i+1}+h\sum_{i=0}^{N_p-1}\sigma_{i}^{x},  \label{TFIM}
\end{equation}
where $\sigma_{i}^{x}$ and $\sigma_{i}^{z}$ are the Pauli matrices, $J$ represents the strength of the spin-spin interaction, and  $h$ is the strength of the external field. 

We use one ancillary qubit to calculate the ground and the first excited state energies of the TFIMs with three and five spin sites. The circuit of AEVQE with one ancillary qubit and five physical qubits is shown in  Fig.\ref{Fig3}(a) on the left side, with the decompositions of the matrices $U$, $W$, and $T$ shown on the right. In the calculation, the ancillary qubits and the physical qubits are initialized to the $(|0\rangle-|1\rangle)/\sqrt{2}$ superposition state. The control-Z (CZ) gate is applied to create entanglement between the ancillary qubit and its adjacent physical qubit. To generate the variational circuit, we use the Trotter decomposition to split the unitary time evolution operator $e^{-iHt}$ into a product of $M$ discrete evolution operators $(e^{-iH\Delta t})^{M}$, where $\Delta t = t/M$ \cite{heyl2019quantum}. Then the discrete time operators are decomposed into single gates and CNOT gates \cite{smith2019simulating,vatan2004optimal}. The optimization results of the loss function with $h/|J|=0.4$ for the three- and five-spin TFIMs are shown in Fig.~\ref{Fig3}(b).

\begin{figure*}
	\centering
	\includegraphics[width=0.99\textwidth]{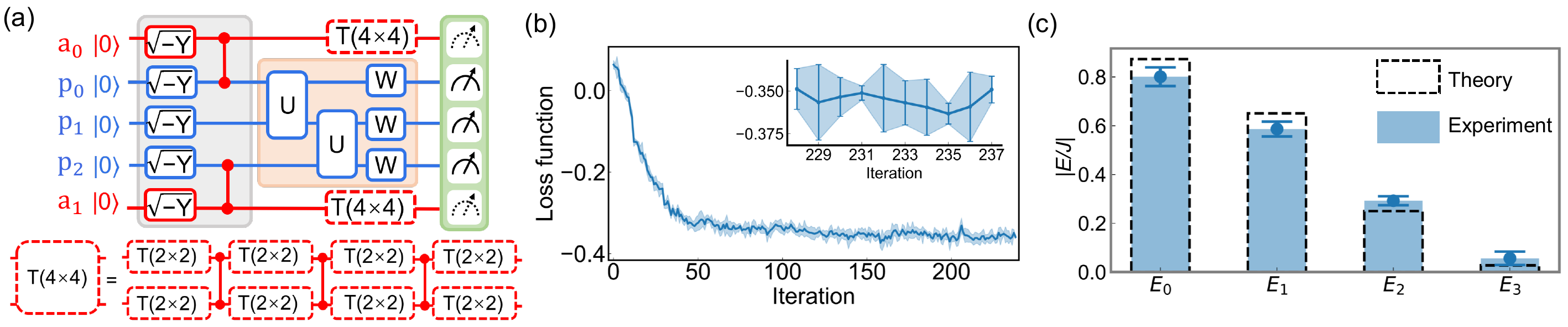}
	\caption{Calculation of the lowest four eigenenergies $E_0$, $E_1$, $E_2$, and $E_3$ of the transverse field Ising model with two ancillary qubits. (a) The circuit schematic for the three-spin system. The transformation matrix $T(4\times4)$ is decomposed into single qubit rotations and CZ gates, as shown at the bottom. The unitary operators $U$, $W$, and $T(2\times2)$ are the same as those in Fig.~\ref{Fig3}(a). (b) The optimization of the loss function for $h/|J| = 0.5$. (c) The experimental eigenenergies $E_0$, $E_1$, $E_2$, and $E_3$ of the three-spin TFIM with $h/|J| = 0.5$. The dashed-line rectangles are the corresponding theoretical results. The error bars represent the standard errors of the average energy.}
	\label{Fig4}
\end{figure*}

\begin{table*}[htbp]
	\centering
	\caption{The observables for  the matrix elements of $H_{\text{sub}}$ for the calculation of the ground and first three excited state energies of the TFIM.}
	\setlength{\tabcolsep}{20pt}
		\begin{tabular}{c|c||c|c}
			\toprule
			Observable   &  Matrix element &  Observable   &  Matrix element\\\hline
			$(I+Z)\otimes (I+Z)  \otimes H$    & $H_{\text{sub}}[0,0]$ &  $(X-iY)\otimes (I+Z) \otimes H$ & $H_{\text{sub}}[2,0]$ \\
			$(I+Z)\otimes (X+iY) \otimes H$    & $H_{\text{sub}}[0,1]$ & $(X-iY)\otimes (X+iY) \otimes H$ & $H_{\text{sub}}[2,0]$ \\
			$(X+iY)\otimes(I+Z)  \otimes H$    & $H_{\text{sub}}[0,2]$ & $(I-Z)\otimes(I+Z) \otimes H$  & $H_{\text{sub}}[2,2]$\\
			$(X+iY)\otimes(X+iY) \otimes H$    & $H_{\text{sub}}[0,3]$ & $(I-Z)\otimes(X+iY)  \otimes H$  & $H_{\text{sub}}[2,3]$ \\
			$(I+Z)\otimes (X-iY) \otimes H$    & $H_{\text{sub}}[1,0]$ &  $(X-iY)\otimes (X-iY) \otimes H$ & $H_{\text{sub}}[3,0]$ \\
			$(I+Z)\otimes (I-Z)  \otimes H$    & $H_{\text{sub}}[1,1]$ &  $(X-iY)\otimes (I-Z)  \otimes H$  & $H_{\text{sub}}[3,1]$ \\
			$(X+iY)\otimes(X-iY) \otimes H$    & $H_{\text{sub}}[1,2]$ & $(I-Z)\otimes(X-iY) \otimes H$  & $H_{\text{sub}}[3,2]$\\
			$(X+iY) \otimes(I-Z)  \otimes H$   & $H_{\text{sub}}[1,3]$ &  $(I-Z)\otimes(I-Z)  \otimes H$ & $H_{\text{sub}}[3,3]$ \\\hline
			\toprule
		\end{tabular}
		\label{table2}
	\end{table*}

After the convergence of the loss function, we use the optimized parameters to calculate the eigenenergies. It is noteworthy that the Hamiltonian $H$ in Eq.~(\ref{TFIM}) commutes with the operator $X = \bigotimes_{i =0}^{N_p-1}\sigma_x^{i}$ ($N_{p}$ = 3, 5), while the ground state $|E_0\rangle$ and the first excited state $|E_1\rangle $ are the eigenstates of $X$ with eigenvalues of +1 and -1, respectively. Hence, symmetry verification can be applied to reduce the errors \cite{cai2023quantum,bonet2018low}.  The symmetry-verified eigenenergies $E_{0,1}$ and density matrices $\rho_{0,1}$ can be calculated according to
\begin{eqnarray}
E_{0,1} = \frac{\text{Tr}[HP_{\pm}\rho P_{\pm}]}{\text{Tr}[P_{\pm}\rho P_{\pm}]},\ \     \rho_{0,1} = \frac{P_{\pm}\rho P_{\pm}}{\text{Tr}[P_{\pm}\rho P_{\pm}]},
\end{eqnarray} 
where $P_{\pm} =  2^{-N_p}(1\pm \bigotimes_{i =1}^{N_p}\sigma_x^{i})$ is the projector. The experimental eigenenergies obtained after  symmetry verification are shown in Figs. \ref{Fig3}(c) and (d). The average energy errors for the three- and five-spin TFIMs are 0.0205 and 0.0987, respectively, showing an increasing error with increasing spin sites.

With the experimentally obtained eigenenergies and eigenstates, we are able to calculate the average absolute magnetization along the Ising chain direction, $m_{\rm z}$, expressed as 
\begin{equation}
m_{\rm z} = \frac{1}{N_p}\sum_{s=0}^{N_p}|N_p-2s|P(s),
\end{equation}
where $P(s)$ represents the probability distribution of the number of spins in state $|1\rangle$, $s = 0,1,...,N_p$. Fig. \ref{Fig4}(e) shows the average absolute magnetization for the three- and five-spin TFIMs as a function of $h/|J|$. The results show an increased steepness with the increase of spin sites, which can be understood considering that for a long transverse Ising chain with infinite spin sites, antiferromagnetic to paramagnetic phase transition would occur at $h/|J|=0.5$.

We next calculate the lowest four eigenenergies of the three-spin TFIM with $h/|J| = 0.5$ by introducing two ancillary qubits. As illustrated in Fig. \ref{Fig4}(a), two pairs of ancillary and physical qubits are prepared in entangled states in the initialization step, while the variational circuit has the same structure as described in Fig. \ref{Fig3}(a). The convergence of the loss function can be seen in Fig. \ref{Fig4}(b). Compared to the experimental result of three-spin TFIM with one ancillary qubit, while the number of optimization parameters remain the same, the average number of iterations nearly double when one additional ancillary qubit is introduced. This suggests that errors from the ancillary qubits may affect the efficiency of  the classical optimizer. After the convergence of the loss function, we measure both the physical and ancillary qubits to calculate $H_\text{sub}$. The observables for the matrix elements of $H_\text{sub}$ are shown in Table \ref{table2}. The eigenenergies obtained by diagonalizing $H_\text{sub}$ are shown in Fig. \ref{Fig4}(c). The errors of the experimental eigenenergies $E_0, E_1, E_2 $ and $E_3$ are 0.072,  0.064, -0.042, and -0.029, respectively. It is important to note that the calculational accuracy does not decrease for higher energy levels, demonstrating the feasibility and advantage of AEVQE in calculating higher excited states. 

\section{Discussion}

\begin{figure}[t]
	\centering
	\includegraphics[width=0.49\textwidth]{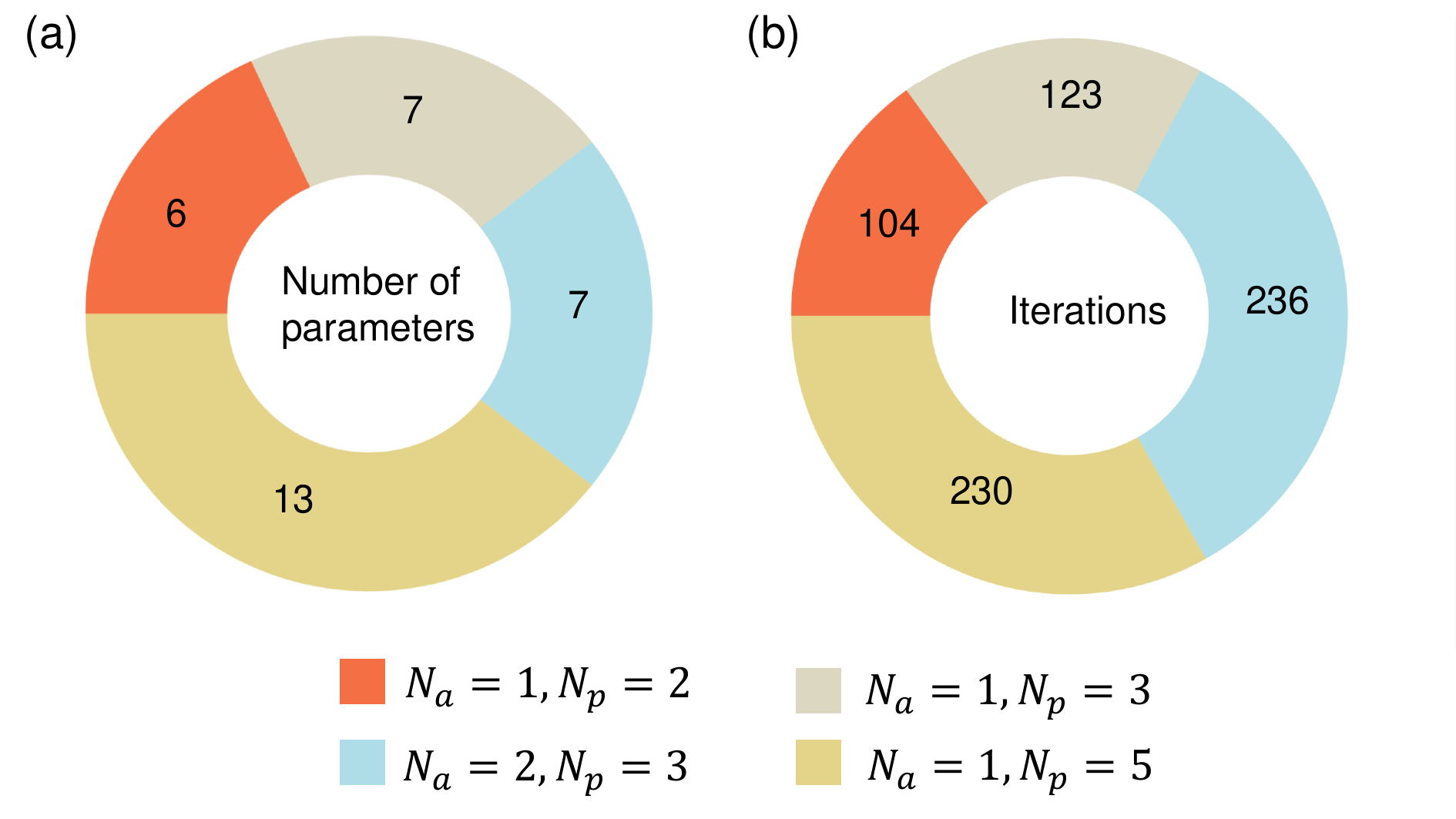}
	\caption{The number of optimization parameters (a) and the average number of iterations (b) for several different experiments indicated. }
	\label{Fig5}
\end{figure}

To compare AEVQE with other algorithms, we illustrate the number of optimization parameters and the average number of iteration steps in Figs.~\ref{Fig5} (a) and (b), respectively, for several experiments indicated. We see that compared to SSVQE in Ref.~\cite{han2024multilevel}, the number of optimization parameters do not depend on the number of ancillary qubits and are the same for the experiments with the same number of physical qubits. In AEVQE, orthogonality is automatically ensured and the orthogonal initial states or ansatz circuits are not required. These significantly reduce the complexity of the quantum circuit design. Also, as the physical system scales up, an increasing number of iteration steps are required. AEVQE increases the optimization efficiency by a factor of $K$ for the calculation of $K$ eigenstates. This is particularly useful in scenarios requiring multiple VQE results, such as in the calculation of the system potential energy curves and the exploration of phase transitions.

Our results also bring out the potential challenges of AEVQE. Firstly, errors on the ancillary qubits may reduce the efficiency of the classical optimizer. This is evidenced by the three-spin TFIM simulations using one and two ancillary qubits. Although the number of optimization parameters remain unchanged, the average number of iterations nearly double when an additional ancillary qubit is introduced. Furthermore, as the number of ancillary qubits increase, creating entanglement between them and the physical qubits becomes more complex. To mitigate initialization errors, higher gate fidelities and qubit connectivity are essential.

In the present experiment, the ancillary qubits are initialized to the superposition state $(|0\rangle \pm |1\rangle)/\sqrt{2}$, ensuring that all eigenenergies contribute equally to the loss function calculation. More generally, the ancillary qubits can be prepared in arbitrary superposition states $\alpha_{i}|0\rangle + \beta_{i}|1\rangle$, where the weights for the calculation of the loss function are determined by product of the coefficients $\alpha_{i},\beta_{i}$ $(i = 0,1,...,N) $. After purification, the eigenstates with lower energies are embedded in the subspace with larger weights. As demonstrated in the numerical study in Ref.~\cite{hong2024Refining}, states with smaller weights tend to exhibit higher error rates. It would be interesting and important to investigate the role of weight in AEVQE through experimental studies. 

\section{Conclusion} 

We have employed AEVQE to simultaneously calculate multiple eigenenergies and eigenstates of the $\text{H}_2$ molecule as well as the transverse field Ising models on a superconducting quantum processor. Different from the previous algorithms, AEVQE does not have the need for multiple orthogonal initialization states or ansatz circuits, thereby improving optimization efficiency by a factor of $K$ when computing $K$ eigenenergies and eigenstates. This advantage is particularly valuable in the applications requiring multiple VQE results, such as calculating the system potential energy curves or exploring quantum phase transitions. Our experiments also demonstrate that AEVQE has high efficiency and accuracy in calculations and requires less computational resources.

\section{Acknowledgments}

We acknowledge supports from the National Natural Science Foundation of China (Grants Nos. 92365206,12404560), the Innovation Program for Quantum Science and Technology (Grant No. 2021ZD0301802).

\section{Data availability}

The data generated in this study have been deposited in the zenodo database \cite{zenodo}.

\end{document}